# Inertial tolerancing and capability indices in an assembly production

**Pierre-Antoine Adragna – Maurice Pillet – Fabien Formosa – Serge Samper**

*Laboratoire SYMME*
*Université de Savoie – Polytech'Savoie*
*B.P. 806*
*F - 74016 Annecy Cedex*
*pierre-antoine.adragna@univ-savoie.fr*
*maurice.pillet@univ-savoie.fr*
*fabien.formosa@univ-savoie.fr*
*serge.samper@univ-savoie.fr*

*RÉSUMÉ. Le tolérancement traditionnel considère la conformité d'un lot lorsque le lot de production satisfait les spécifications. La caractéristique est considérée pour elle-même mais pas en fonction de son incidence dans l'assemblage dont elle fait partie. Le tolérancement inertiel propose une autre alternative de tolérancement afin de garantir l'assemblage final. L'inertie* $I=\sqrt{\delta^2+\sigma^2}$ *n'est pas tolérancée par un intervalle de tolérance mais par un scalaire représentant l'inertie maximale que la caractéristique ne doit pas dépasser. Nous détaillerons comment calculer ces tolérances inertielles selon deux cas de figures, si on vise à respecter une inertie ou un indice de capabilité Cpk sur la caractéristique résultante dans le cas particulier de tolérances uniformes et le cas général de tolérances non uniformes. Un exemple servira à comparer les différentes méthodes de tolérancement.*

*ABSTRACT. Traditional tolerancing considers the conformity of a batch when the batch satisfies the specifications. The characteristic is considered for itself and not according to its incidence in the assembly. Inertial tolerancing proposes another alternative of tolerancing in order to guarantee the final assembly characteristic. The inertia* $I=\sqrt{\sigma^2+\delta^2}$ *is not toleranced by a tolerance interval but by a scalar representing the maximum inertia that the characteristic should not exceed. We detail how to calculate the inertial tolerances according to two cases, one aims to guarantee an inertia of the assembly characteristic the other a tolerance interval on the assembly characteristic by a Cpk capability index, in the particular but common case of uniform tolerances or more general with non uniform tolerances. An example will be detailed to show the results of the different tolerancing methods.*

## 1. Introduction

The traditional tolerancing considers the conformity of a batch when the batch satisfies the specifications. The characteristic is considered for itself and not regarding its incidence on the final assembly resultant. It has been showed that inertial tolerancing ($I = \sqrt{\sigma^2 + \delta^2}$ which is no more based on a *[Min Max]* interval but on the Taguchi loss function) proposes another tolerancing method to guarantee the final assembly while allowing larger variability in the case of centered production.

This paper proposes a method to calculate the inertial tolerances of the components of a 1D mechanical assembly chain in two cases:

- we want to guarantee an inertial tolerance on the final assembly,

- we want to guarantee a minimum of the *Cpk* index on the tolerance interval *[Min; Max]* of the assembly characteristic.

Different cases are considered, even the general case where the components tolerances are not uniformly distributed. A comparison with the traditional tolerancing will show the difference on the allowed variability of components. The two considered cases have different hypothesis of application. We will discuss on the choice of using the first or the second approach.

An industrial case of application will be treated as an example.

## 2. Inertial tolerancing, capability indices and conformity

The aim of tolerancing is to determine an acceptation criterion on the components characteristics $x_i$ to guarantee the quality of the assembly resultant $Y$. In the case of a good design, when the $x$ characteristic is produced on the target, the quality is optimal. As $x$ gets an offset from the target, the function of the assembly will be more sensitive to the conditions of use and the environment, and can lead to a non-satisfaction of the customer. Taguchi demonstrated that the financial loss associated to an offset from the target is proportional to the square of this off-centering $L = K(T-X)^2$ . (Pillet *et al*, 2001) shows that in the case of a batch, the financial loss associated is $L = K(\sigma^2 + \delta^2)$. Then he defines

$$I_x = \sqrt{\frac{\sum_{i=1}^{n}(x - Target)^2}{n}} = \sqrt{\sigma_x^2 + \delta_x^2} \qquad [1]$$

This function is called Inertia by analogy to the mechanical inertia. Here $I_x$ represents the inertia of the $x$ characteristic, $\sigma_x$ is the standard deviation of the batch distribution and $\delta_x$ is the offset of the mean to the target.

To qualify the capability of a process with the inertial tolerancing, two capability indices have been defined:

$$Cp = \frac{I_{Max}}{\sigma_{Batch}} \quad [2]$$

which indicates the capability of a centered process.

$$Cpi = \frac{I_{Max}}{I_{Batch}} \quad [3]$$

which indicates the capability considering the process off-centering.

Compared to the traditional tolerancing, the proposed approach of the inertial tolerancing is quite different. The aim is no more to guarantee a rate of parts out of tolerance, but to guarantee the centering of components around the target in order to guarantee the quality of the assembly. The reflection is no more based on the proportions out of tolerances but on the inertias of the components, the normality of the batch distribution is no more a necessary criterion.

## 3. Tolerancing of an assembly

For the tolerancing of assembly systems, the problem consists of finding the elementary characteristics $x_i$ of the components in order to obtain a final characteristic $Y$ satisfying the functional condition of the assembled product for the customers needs. As a general rule, it is possible to approximate the system behavior around the target by a linearization at the first order. The final characteristic behavior can be expressed by the following relation:

$$Y = \alpha_0 + \sum_{i=1}^{n} \alpha_i . x_i \quad [4]$$

Where $\alpha_i$ is the influence coefficient of the component $i$ on the assembly resultant $Y$, $\alpha_0$ is the target value of $Y$ and $n$ is the number of components in the assembly. For the computation of the components tolerances, the general case will be considered where tolerances are not uniformly distributed with the use a difficulty coefficient $\beta_i \geq 1$, also called feasibility coefficient. The simplified but common case will also be presented, where the tolerances are uniformly distributed $\beta_i = 1$, and the incidence coefficients are all equal $\alpha_i = 1$ .

### 3.1. *Review of the traditional tolerancing methods*

Before the presentation of the inertial tolerancing, here is a brief reminds of the traditional tolerancing methods of assembly systems. Three commonly used traditional methods are presented.

#### *3.1.1. Worst of cases tolerancing*

In this case, one considers that the final characteristic of the assembly will be respected in any cases of assembly. In the general case where tolerances are non-uniformly distributed, the $\beta_i$ coefficients will be used for the components. The tolerance expression of a components is $R_{xi} = \beta_i \,.\, R_x$ , the assembly resultant is then:

$$R_Y = \sum_{i=1}^{n} \left( |\alpha_i| \,.\, \beta_i \,.\, R_x \right) \qquad [5]$$

$$R_{xi} = \beta_i \,.\, \frac{R_Y}{\sum_{i=1}^{n} \left( |\alpha_i| \,.\, \beta_i \right)} \qquad [6]$$

Where $R_Y$ represents the tolerance interval of the functional condition of the assembly, and $R_{xi}$ is the tolerance interval of the components. Tolerances can be distributed regarding different methods by changing the $\beta_i$ coefficients (Graves, 2001):

- Uniform distribution of the tolerance,
- Considering the tolerances of standard components, or conception rules,
- Proportional to the square of the nominal length,
- Considering the process capabilities,

In the case of uniform distribution of the tolerance $\beta_i = 1$ , and same incidences of the components $\alpha_i = 1$ , the relation [5] becomes then:

$$R_Y = \sum_{i=1}^{n} R_{xi} = n \, . \, R_{xi} \qquad [7]$$

$$R_{xi} = \frac{R_Y}{n} \qquad [8]$$

The well known inconvenient of this method is the high price of production due to the tightened tolerances of the components.

3.1.2. *Statistical tolerancing*

The statistical tolerancing has been developed to consider the low probability of having several characteristics in limit of their tolerances simultaneously (Shewhart, 1931) (Evans, 1975). Under the hypothesis that the $x_i$ variables are independent with a standard deviation $R_{xi}$, the equation [4] gives the following relation:

$$\sigma_Y = \sqrt{\sum_{i=1}^{n} (\alpha_i^2 \, . \, \sigma_{xi}^2)} \qquad [9]$$

In the general case of non uniform distribution, one consider $R_{xi} = \beta_i \, . \, \sigma_{xi}$ , then:

$$R_Y = \sqrt{\sum_{i=1}^{n} (\alpha_i^2 \, . \, \beta_i^2 \, . \, \sigma_{xi}^2)} \qquad [10]$$

$$R_{xi} = \beta_i \, . \, \frac{R_Y}{\sqrt{\sum_{i=1}^{n} (\alpha_i^2 \, . \, \beta_i^2)}} \qquad [11]$$

With tolerances proportional to the standard deviation (Chase *et al.,* 1991), in the case of a uniform repartition, and same incidence of the components $\alpha_i = 1$ , the relation [10] becomes then:

$$R_Y = \sqrt{\sum_{i=1}^{n} R_{xi}^2} = \sqrt{n \, . \, R_{xi}^2} \qquad [12]$$

$$R_{xi} = \frac{R_Y}{\sqrt{n}} \qquad [13]$$

In this tolerancing method, the basic hypothesis is to consider the centering of all characteristics on their target values. The inconvenient of this method is that it does not guarantee the conformity of the assembly characteristic in all configurations of the components. It can be possible that the components respect their tolerance intervals, but their off-centering from the target lead to non-conformity on the final condition in its tolerance interval.

3.1.3. *Inflated statistical tolerancing:*

Several methods are proposed in order to reduce the negative aspect of the statistical tolerancing. A proposed method is the inflated statistical tolerancing (Graves, 1997) (Graves, 2001). This method consists of using the inflated coefficient in the tolerancing of the components based on the statistical method.

$$R_Y = f \, . \sqrt{\sum_{i=1}^{n} (\alpha_i^2 \, . \, \beta_i^2 \, . \, R_{xi}^2)} \qquad [14]$$

$$R_{xi} = \beta_i \, . \, \frac{R_Y}{f \, . \sqrt{\sum_{i=1}^{n} (\alpha_i^2 \, . \, \beta_i^2)}} \qquad [15]$$

Where $f$ represents the inflated coefficient generally chosen around $f = 1,5$ to $1,6$. In the case of a uniform distribution of the tolerances and same incidences of the components, one has the following relation:

$$R_Y = f \, . \sqrt{\sum_{i=1}^{n} R_{xi}^2} \qquad [16]$$

$$R_{xi} = \frac{R_Y}{f \, . \sqrt{n}} \qquad [17]$$

In the case where $f = 1$, one finds the statistical tolerancing method. In the case where $f = \sqrt{n}$, one finds the worst of cases tolerancing method. A discussion (Graves *et al*, 2000) on different situations leads to choose different $f$ values. Graves proposes an interesting approach to optimize the $f$ coefficient following different capability indices. Although this is an improved method, it is possible to find a situation where the final assembly characteristic will not well be respected.

### 3.2. *Inertial tolerancing of an assembly guarantying an inertia on the resultant*

The following results come from (Pillet, 2002). These results are reminded in order to be used further. The calculations are done under the hypothesis of a uniform distribution of the components tolerances.

$$\sigma_Y = \sqrt{\sum_{i=1}^{n} \alpha_i^2 \, . \, \sigma_{xi}^2} \qquad [18]$$

$$\delta_Y = \sum_{i=1}^{n} \alpha_i \, . \, \delta_{xi} \qquad [19]$$

The inertia of the resultant characteristic $Y$ is defined by the relation [1]. Replacing relations [18] and [19], one have the inertia of the resultant in function of the components inertias and off-centering:

$$I_Y = \sqrt{\sum_{i=1}^{n} \alpha_i^2 \,.\, I_{xi}^2 + 2 \,.\, \sum_{i=1}^{n} \alpha_i \,.\, \alpha_j \,.\, \delta_{xi} \,.\, \delta_{xj}} \qquad [20]$$

The first part of the equation corresponds to the addition of the squared inertias. The double product corresponds to the case where all off-centering of the components are on the same side. A discussion is necessary to treat different hypothesis.

### *3.2.1. Hypothesis 1: Worst of cases*

This hypothesis of components in their worst of cases considers that the component inertia is only due to its off-centering from the target, $I_{xi} = \sqrt{\delta_{xi}^2} = |\delta_{xi}|$, the relation [20] becomes then:

$$I_Y = \sqrt{\sum_{i=1}^{n} \alpha_i^2 \,.\, \delta_{xi}^2 + 2 \,.\, \sum_{i=1}^{n} \alpha_i \,.\, \alpha_j \,.\, \delta_{xi} \,.\, \delta_{xj}} \qquad [21]$$

In the case where all the $\alpha_i = 1$, and the tolerances are uniformly distributed $I_{xi} = \delta_{xi}$, one have the following relations:

$$I_Y = \sqrt{n^2 \,.\, I_x^2} = n \,.\, I_x \qquad [22]$$

$$\boxed{I_x = \frac{I_Y}{n}} \qquad [23]$$

In the general case where the $\alpha_i$ coefficients are non-equal and the tolerances are not uniformly distributed $I_{xi} = \beta_i \,.\, \delta_x$, one has the following relation:

$$I_{xi} = \beta_i \,.\, \frac{I_Y}{\sum_{i=1}^{n} |\alpha_i| \,.\, \beta_i} \qquad [24]$$

*3.2.2. Hypothesis 2: random distribution of the averages*

This hypothesis is close to the consideration made for the traditional statistical tolerancing. In this case, the double product $\sum_{i=1}^{n} \alpha_i \,.\, \alpha_j \,.\, \delta_{xi} \,.\, \delta_{xj}$ is null, the relation [20] becomes then:

$$I_Y = \sqrt{\sum_{i=1}^{n} \alpha_i^2 \,.\, I_{xi}^2} \qquad [25]$$

In the case where $\alpha_i = 1$, one thus have:

$$I_Y = \sqrt{n \,.\, I_x^2} \qquad [26]$$

$$\boxed{I_x = \frac{I_Y}{\sqrt{n}}} \qquad [27]$$

In the general case of non-uniform distribution of tolerances, one considers $I_{xi} = \beta_i \,.\, \sigma_{xi}$ . From relation [25] one obtains:

$$I_Y = \sqrt{\sum_{i=1}^{n} \left( \alpha_i^2 \,.\, \beta_i^2 \,.\, \sigma_{xi}^2 \right)} \qquad [28]$$

thus the components inertia can be

$$I_{xi} = \beta_i \,.\, \frac{I_Y}{\sqrt{\sum_{i=1}^{n} (\alpha_i^2 \,.\, \beta_i^2)}} \qquad [29]$$

*3.2.3. Hypothesis 3: off-centering of* $\boldsymbol{\delta = k \,.\, \sigma}$ *of all components.*

This hypothesis considers that all components have systematic off-centering equals to $\delta = k \,.\, \sigma$. In this case the component inertia is:

$$I_x = \sqrt{\sigma_x^2 + (k \,.\, \sigma_x)^2} = \sigma_x \,.\, \sqrt{1 + k^2} \qquad [30]$$

then

$$\sigma_x = \frac{I_x}{\sqrt{1 + k^2}} \qquad [31]$$

and

$$\delta_x = I_x \,.\, \sqrt{1 - \frac{1}{1 + k^2}} \qquad [32]$$

Equation [20] becomes then:

$$I_Y = \sqrt{\sum_{i=1}^{n} \alpha_i^2 \,.\, I_{xi}^2 + 2 \,.\, \sum_{i=1}^{n} \alpha_i \,.\, \alpha_j \,.\, \left(1 - \frac{1}{1 + k^2}\right) .\, I_{xi} \,.\, I_{xj}} \qquad [33]$$

In the case where all $\alpha_i = 1$, one obtains then:

$$I_Y = \sqrt{n.I_x^2 + n.(n-1).\left(1 - \frac{1}{1+k^2}\right).I_x^2} \qquad [34]$$

thus

$$\boxed{I_x = \frac{I_Y}{n.\left(\frac{n.k^2+1}{1+k^2}\right)}} \qquad [35]$$

### *3.2.4. Hypothesis 4: off-centering of **m** components out of **n***

In this hypothesis, the designer determines the number of characteristics that can have a systematic off-centering. Under these conditions, the double product is reduced. In the case where all $\alpha_i = 1$ and tolerances are uniformly distributed, one have:

$$I_Y = \sqrt{n.I_x^2 + m.(m-1).\left(1 - \frac{1}{1+k^2}\right).I_x^2} \qquad [36]$$

thus

$$\boxed{I_x = \frac{I_Y.\left(1+k^2\right)}{n.\left(1+k^2\right) + m.k^2.(m-1)}} \qquad [37]$$

## 4. Inertial tolerancing guarantying a *Cpk* index on the assembly resultant: the corrected inertial tolerancing

In the case where the aim is to guarantee a *Cpk* index on the assembly resultant, the conformity of this final characteristic is considered regarding a tolerance interval. In the most common case where the target is centered in the tolerance interval $Target = \frac{LT + UT}{2}$ where $LT$ is the lower tolerance and $UT$ is the upper tolerance and $R_Y = \frac{UT - LT}{2}$ is the length of the tolerance interval. The capability index *Cpk* is defined by:

$$Cpk = Min\left(\frac{\frac{R_Y}{2} - \left(Target - \overline{x}\right)}{3.\sigma_Y}, \frac{\frac{R_Y}{2} - \left(\overline{x} - Target\right)}{3.\sigma_Y}\right) \quad [38]$$

Which is equivalent to

$$Cpk = \frac{\frac{R_Y}{2} - \left|\delta_Y\right|}{3.\sigma_Y} \quad [39]$$

Replacing with the equations [18] and [19], one has the following relation:

$$Cpk = \frac{\frac{R_Y}{2} - \left|\sum_{i=1}^{n} \alpha_i . \delta_{xi}\right|}{3 . \sqrt{\sum_{i=1}^{n} \alpha_i^2 . \sigma_{xi}^2}} \quad [40]$$

In order to define the components inertia in function of the tolerance interval to guarantee on the assembly resultant, one considers that the components are centered. In this case, the components inertia is only due to their standard deviation: $I_{xi} = \sqrt{\sigma_{xi}^2 + 0} = \sigma_{xi}$ . The resultant is also centered, from where $I_Y = \sigma_Y$ , one have then:

$$Cp = \frac{R_Y}{6 . \sigma_Y} \quad [41]$$

$$\sigma_Y = I_Y = \frac{R_Y}{6 . Cp} \quad [42]$$

In the general case of non-uniform distribution, one considers $I_{xi} = \beta_i \,.\, \sigma_{xi}$ . In order to differentiate the *Cp* index to a prediction index, it will be renamed as the *ICC* coefficient (Inertial Corrected Coefficient). As the hypothesis of centered production is considered, one has $I_Y = \sigma_Y$ . From the relation [29] one obtains the following tolerancing relation:

$$I_{xi} = \beta_i \,.\, \frac{R_Y}{6 \,.\, ICC \,.\, \sqrt{\sum_{i=1}^{n} (\alpha_i^2 \,.\, \beta_i^2)}} \qquad [43]$$

In the case of a uniform repartition of the tolerances and all $\alpha_i = 1$, one obtains then:

$$I_{xi} = \frac{R_Y}{6 \,.\, ICC \,.\, \sqrt{n}} \qquad [44]$$

#### *4. 1. Variations of the Cpk index in function of the components off-centering*

Now it is possible to define the components inertia in function of the tolerance interval of the assembly resultant, let us interest in the variations of the *Cpk* index while components are in limit of their inertial tolerances. This hypothesis allows expressing the standard deviation in function of the off-centering, one has the expression of the batch inertia given by [1], $I_{xi} = \sqrt{\sigma_{xi}^2 + \delta_{xi}^2}$ thus the components deviation can be expressed as follows:

$$\sigma_{xi} = \sqrt{I_{xi}^2 - \delta_{xi}^2} \qquad [45]$$

Which is replaced in [39]. Let us also consider that all $\delta_{xi} = \delta$, one obtains then:

$$Cpk = \frac{\frac{R_Y}{2} - n.\left|\delta\right|}{3 \,.\, \sqrt{n.\left(I_{xi}^2 - \delta^2\right)}} \qquad [46]$$

The component inertias are defined by [44], then:

$$Cpk = \frac{\frac{R_Y}{2} - n.|\delta|}{3\,.\sqrt{\frac{R_Y^{\,2}}{36\,.\,ICC^2} - n.\delta^2}} \qquad [47]$$

The *Cpk* function is pair, the domain of study can then be reduced to $\delta > 0$ , this allows to suppress the absolute value function on $\delta$. The derivate function of *Cpk* at $\delta$ is:

$$\frac{\partial Cpk}{\partial \delta} = \frac{\left(\frac{R_Y}{2} - n.\,\delta\right)n.\delta}{3\,.\left(\frac{R_Y^{\,2}}{36\,.\,ICC^2} - n.\delta^2\right)^{3/2}} - \frac{n}{3\,.\sqrt{\frac{R_Y^{\,2}}{36\,.\,ICC^2} - n.\delta^2}} \qquad [48]$$

The variations of the *Cpk* function are studied thanks to the sign of its derivate function continuously defined on $[0; I_{xi}[$. One obtains the following table of variations.

| $\delta$ | $0$ | | $\delta = \frac{R_Y}{18.ICC^2}$ | | $I_{xi} = \frac{R_Y}{6.ICC.\sqrt{n}}$ |
|---|---|---|---|---|---|
| *Cpk* | *ICC* | ↘ | $\sqrt{ICC^2 - \frac{n}{9}}$ | ↗ | *+ Infinity* |
| $\frac{\partial Cpk}{\partial \delta}$ | $-\frac{2.n.ICC^2}{R_Y}$ | – | $0$ | + | *+ Infinity* |

**Table 1.** *Variations of the assembly Cpk function depending on the component off-centering* $\delta$ *for a given number of components toleranced by the corrected inertial tolerancing.*

The function has a minimum for:

$$\delta^{Min} = \frac{R_Y}{18.ICC^2} \quad [49]$$

And the value of *Cpk* is:

$$Cpk^{Min} = \sqrt{ICC^2 - \frac{n}{9}} \quad [50]$$

These results are interesting because they allow to find the minimum of the *ICC* coefficient value to guarantee a minimum of the *Cpk* index in function of the number of components in the assembly. Thus it is also possible to know how many components can compose the assembly in order to guarantee a minimum of the resultant *Cpk* index with a given value of the *ICC* coefficient:

$$ICC = \sqrt{Cpk^2 + \frac{n}{9}} \quad [51]$$

$$n = 9.(ICC^2 - Cpk^2) \quad [52]$$

NOTE. — The results found considering the hypothesis that all component off-centering are equal $\delta_{xi} = \delta$ correspond to those found with the gradient method. It consists of calculating the $\Delta f$ gradient of the *Cpk* function depending on *n* variable $\delta_{xi}$ and its *Hf* Hessian matrix, and then to find the point for which the gradient is null and the eigen values of the Hessian matrix are all strictly positive in order to have a local minimum. In our case, the local minimum of the *Cpk* is found for all $\delta_{xi} = \frac{R_Y}{18.ICC^2}$. A demonstration by recurrence on *n* and *i* proves it but will not be exposed in this paper.

### *4. 2. Variations of the Cpk function regarding the number of components and their off-centering*

We will see the influence of the off-centering and the number of components on the *Cpk* index value. Figure 1 confirms the previous results on the *Cpk* variations. An assembly system which components tolerancing has been done with $ICC = 1$ or $1,5$ has a *Cpk* which is minimum when all $\delta_{xi} = 0,056$ or $\delta_{xi} = 0,025$ respectively.

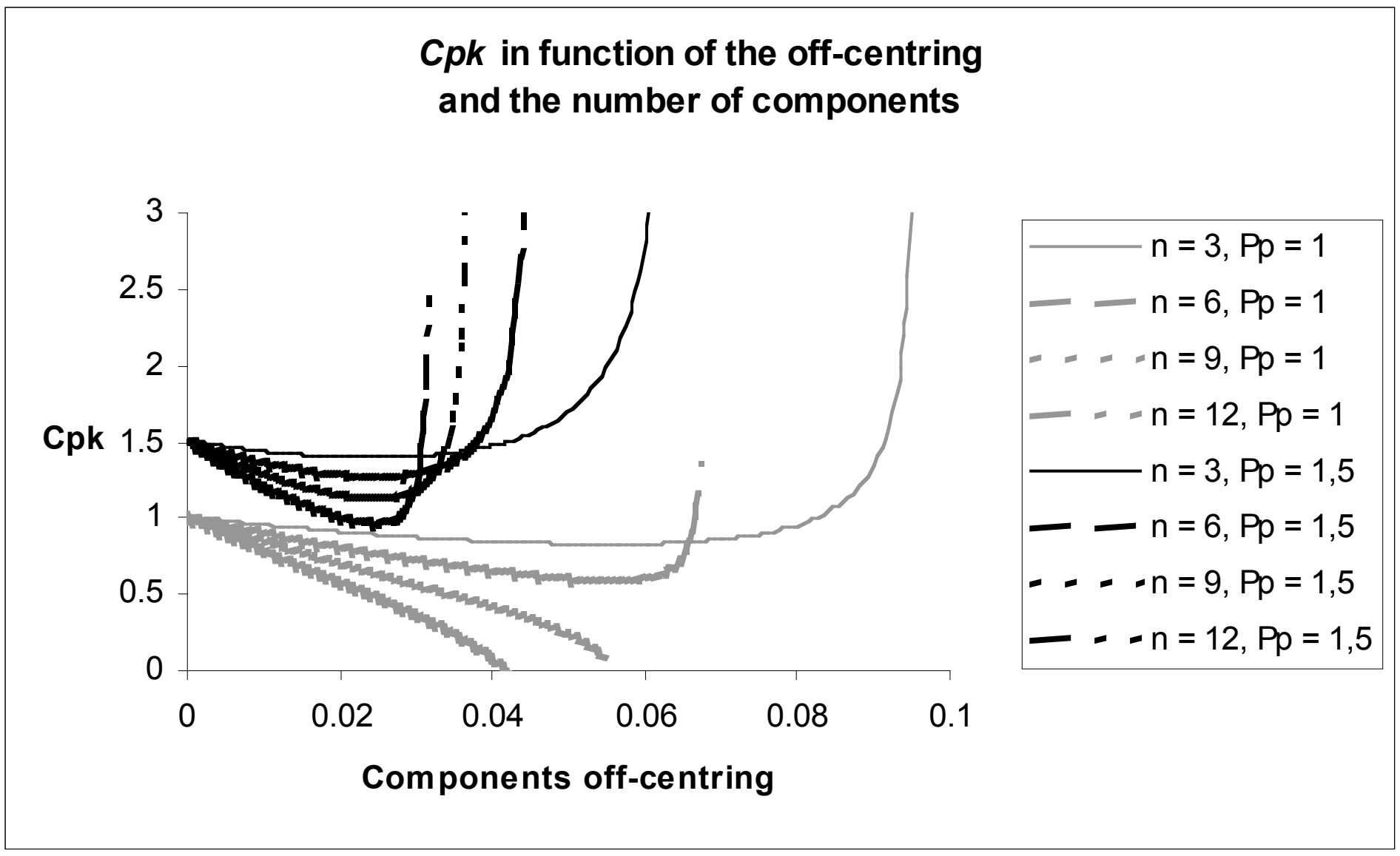

**Figure1.** *Variations of Cpk indices with different configurations of the number of components and ICC coefficient values in function of the components off-centering in limit of their inertial tolerances.*

The Cpk minimum can be calculated for $n < 9.ICC^2$, from relation [50]. This remark can also be observed on the following table where *ICC = 1* and *1,5*, and *n = 1* to *12*. For an *ICC* index chosen to *ICC = 1*, the minimum value of the *Cpk* index can be calculated up to *n = 9*, for *n = 12* it is not possible to evaluate the *Cpk* minimum. With a limit study, it can be showed that this minimum tends toward - *infinity*.

| *Cpk*, $R_Y = 1$ | *n = 3* | *n = 6* | *n = 9* | *n = 12* |
|---|---|---|---|---|
| *ICC = 1* | *0,816* | *0,577* | *0,000* | *- infinity* |
| *ICC = 1,5* | *1,384* | *1,258* | *1,118* | *0,957* |

**Table 2.** *Different values of the minimum of the Cpk index for different number of components n and different ICC coefficient values.*

A link can be observed between the evolution of the minimum of the *Cpk* index and the *ICC* coefficient value. With an *ICC* coefficient *ICC = 1*, it is impossible to guarantee a *Cpk* index, *Cpk > 1* due to this worst configuration of all $\delta_{xi}$ , even for $n = 1$ when $\delta \neq 0$. For an *ICC* coefficient *ICC = 1,5*, it is possible to guarantee a *Cpk* index $Cpk > 1,1$ up to $n = 9$ components in the assembly. It is nearly possible to guarantee $Cpk = 1$ for $12$ components.

Let us see the variations of the *ICC* coefficient value for the calculation of the inertial tolerances in function of the *Cpk* index value to guarantee, and the number of components in the assembly.

### *4. 3. ICC values to guarantee a minimum of the Cpk index in function of the number of components.*

From relation [51], one can choose the *ICC* coefficient value in order to guarantee a minimum of the *Cpk* index of the assembly in its tolerance interval.

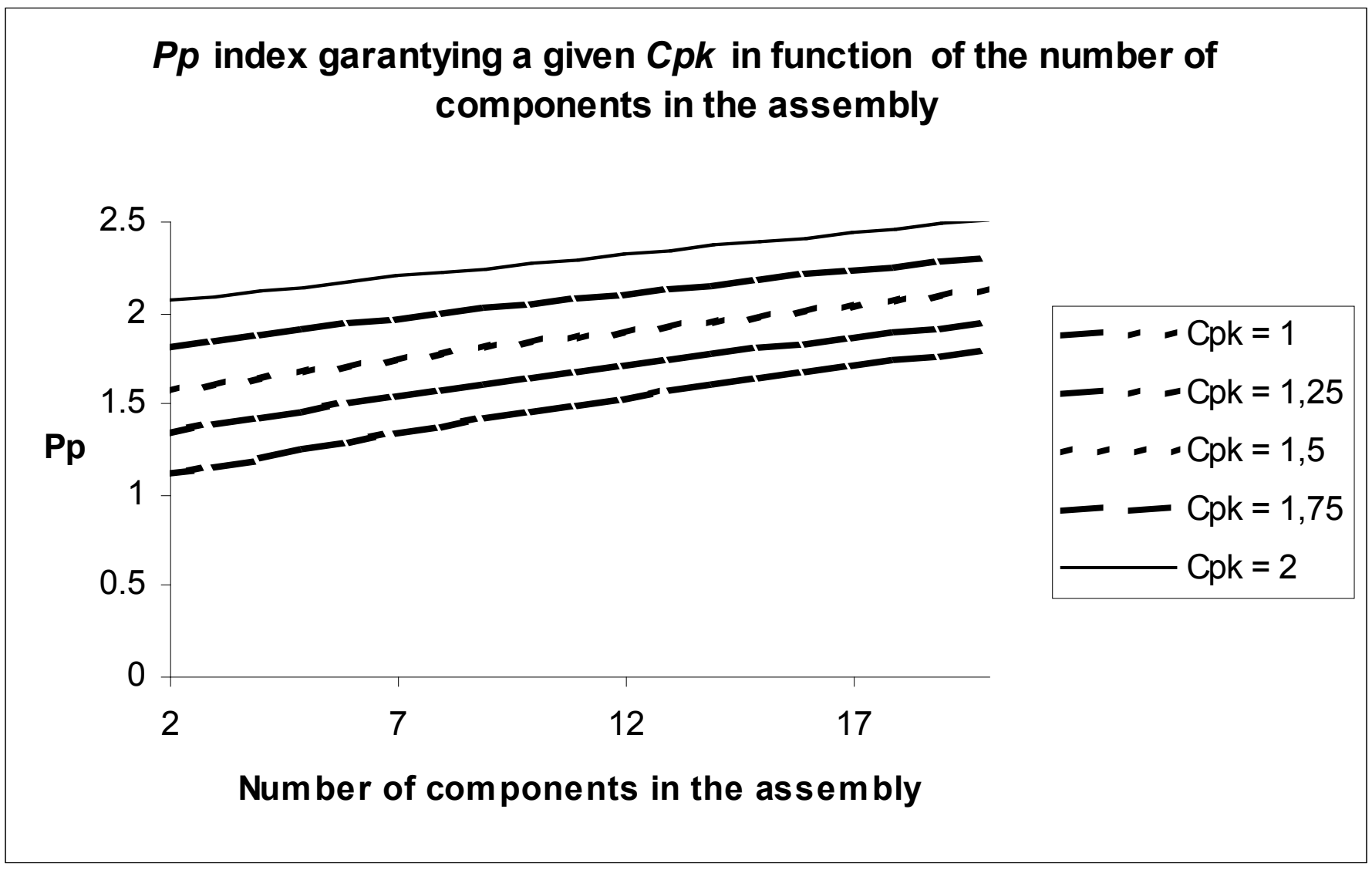

**Figure 2.** *Value of the ICC coefficient to take into account in order to guarantee different values of the Cpk index in function of the number of tolerance.*

Figure 2 is an abacus of the inertial tolerancing guarantying a *Cpk* index on the assembly resultant in the case of a uniform distribution of the tolerances and same incidences of the components $\alpha_i = 1$ . Two relations have to be considered while dealing with the particular but common case of a uniform distribution of the tolerances for an assembly where the components have the same incidences, the relations [44], $I_{xi} = \dfrac{R_Y}{6\,.\,ICC\,.\,\sqrt{n}}$, and [51], $ICC = \sqrt{Cpk^2 + \dfrac{n}{9}}$.

### *4.4. Validation of the abacus in the case of the generalized tolerancing*

Previous results are obtained in the case of a particular inertial tolerancing with uniform distribution of the tolerances and all $\alpha_i = 1$. What are the results in the general case of a non-uniformly distributed tolerancing and all incidences are not equal? In the general case of non-uniform repartition of the tolerances, the inertia is defined by the relation [43]

Replacing [43] and [45] in the equation of the *Cpk* defined by [40], we obtain the following expression of the *Cpk*:

$$Cpk = \frac{\frac{R_Y}{2} - \sum_{i=1}^{n} \alpha_i \, . \, |\delta_{xi}|}{3 \, . \sqrt{\frac{R_Y^2}{36 \, . \, ICC^2} - \sum_{i=1}^{n} \alpha_i^2 \, . \, \delta_{xi}^2}} \qquad [53]$$

One can feel that with a variable change $\delta = \alpha_i \, . \, \delta_{xi}$, the minimum of the *Cpk* can be found for the general case for $\delta_{xi} = \frac{R_Y}{18 \, . \, ICC^2 . \, \alpha_i}$, in this case $Cpk = \sqrt{ICC^2 - \frac{n}{9}}$ as in the particular case. One can conclude that the abacus presented in figure 2 is also applicable in the case of generalized inertial tolerancing.

## 5. An application case

The example will be used to compare the results of the different tolerancing methods.

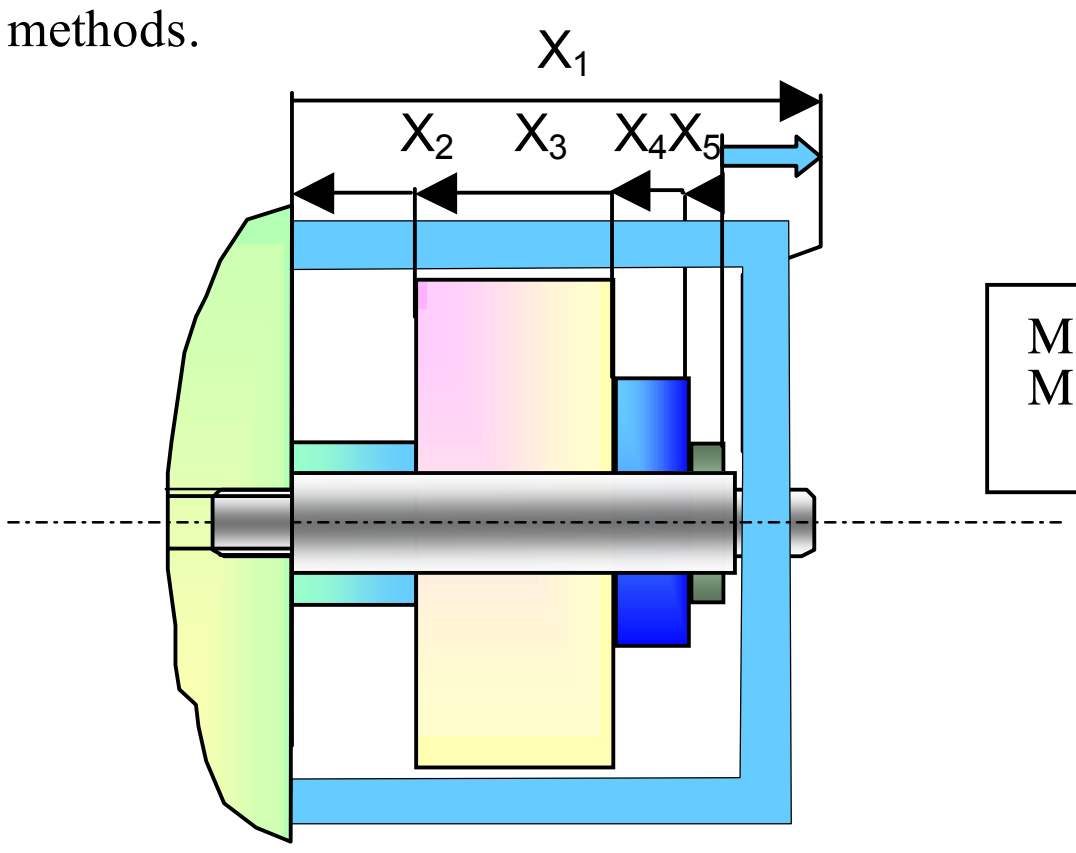

Mini on the condition (LT): 0.5
Maxi on the condition (UT) : 1.5

The target value of the gap is *1mm* with a tolerance interval of *1mm*, thus *+/- 0,5mm*. The tolerance distribution is uniform. The following relation defines the assembly relation:

$$Y = X_1 - X_2 - X_3 - X_4 - X_5 \qquad [54]$$

### 5.1. *Application of the different tolerancing methods*

The inertial tolerancing of the two-presented cases (guarantying an inertia or an interval) will be compared to the traditional tolerancing methods. In the case of application, the tolerance interval $R_Y = 1$, the number of components is $n = 5$. In the case of an inertial tolerancing guarantying a capability index $Cpk = 1$ on the assembly resultant, the *ICC* coefficient to take into account for the inertial tolerances calculation is:

$$ICC = \sqrt{Cpk^2 + \frac{n}{9}} = 1,25 \qquad [55]$$

| ***Traditional tolerancing*** | | ***Inertial tolerancing*** | |
|---|---|---|---|
| *Worst of case:* $R_{xi} = \frac{R_Y}{n}$ | $R_{xi} = 0{,}200$ *thus* $\boldsymbol{\sigma_{xiMax} = 0{,}033}$ | *Guarantying an inertia by worst of case:* $I_{xi} = \frac{R_Y}{6.n}$ | $\boldsymbol{I_{xi} = 0{,}033}$ |
| *Statistical:* $R_{xi} = \frac{R_Y}{\sqrt{n}}$ | $R_{xi} = 0{,}447$ *thus* $\boldsymbol{\sigma_{xiMax} = 0{,}075}$ | *Guarantying an inertia by statistical:* $I_{xi} = \frac{R_Y}{6.\sqrt{n}}$ | $\boldsymbol{I_{xi} = 0{,}075}$ |
| *Inflated statistical, f = 1,5:* $R_{xi} = \frac{R_Y}{1{,}5.\sqrt{n}}$ | $R_{xi} = 0{,}298$ *thus* $\boldsymbol{\sigma_{xiMax} = 0{,}050}$ | *Guarantying a Cpk ≥ 1 with ICC = 1,25:* $I_{xi} = \frac{R_Y}{6.1{,}25.\sqrt{n}}$ | $\boldsymbol{I_{xi} = 0{,}060}$ |

**Table 2.** *Comparison of different traditional and inertial tolerancing methods. For the traditional tolerancing, the tolerance interval of the components is* $R_{xi}$, *and in the case of a centered batch, the maximum batch dispersion is* $\sigma_{xiMax} = \frac{R_{xi}}{6}$. *For the inertial tolerancing, the components inertias are expressed by* $\sigma_{xiMax} = I_{xi}$ .

The maximum batch dispersion in the case of the traditional tolerancing is useful to compare the allowed dispersion on the components with the two methods. For inertial tolerancing, when the batch is centered, the maximum batch dispersion equals to the inertial tolerance $I_{xi}$. It is then possible to compare the maximum dispersion allowed for centered components with the different tolerancing methods.

### 5.2. *Discussion on the case of application*

One can see that the maximum authorized dispersions for the traditional tolerancing by worst of cases or statistical are similar to those of the inertial tolerancing guarantying an inertia by worst of cases $\sigma_{xiMax} = I_{xi} = 0{,}033$, or statistical $\sigma_{xiMax} = I_{xi} = 0{,}075$ respectively. But the maximum allowed dispersion by the inflated statistical tolerancing is $\sigma_{xiMax} = 0{,}05$, which is lower than the maximum authorized dispersion by the corrected inertial tolerancing guarantying a minimum of the *Cpk* index $Cpk = 1$ which is $I_{xi} = 0{,}06$.

Before concluding that this last method guarantying a *Cpk* index gives larger dispersion than the inflated statistical tolerancing, let us compare the influence of the number of components on this dispersion difference. First, the inflated coefficient $f = 1,5$ can be compared to the *ICC* coefficient $ICC = 1,25$ that is lower. Choosing a *ICC* coefficient $ICC = 1,5$ is the same as guarantying a *Cpk* index $Cpk = 1,30$ for an assembly of 5 components, from relation [52], but this also guarantee a *Cpk* index $Cpk = 1$ for an assembly composed of up to 11 components from relation [50], $11,25 = n = 9.(ICC^2 - Cpk^2)$ . It is possible to conclude that for an assembly of less than 12 components, the inertial tolerancing guarantying a *Cpk* index $Cpk = 1$ is larger than the inflated statistical tolerancing. Over *12* components, the inflated statistical tolerancing gives larger dispersions than the inertial tolerancing guarantying a *Cpk* index, but the inertial tolerancing allows to guarantee a *Cpk* index on the final assembly characteristic, that cannot do the inflated statistical tolerancing.

The comparison between traditional statistical tolerancing and inertial tolerancing thanks to the authorized dispersion is dangerous; this comparison can be done only for centered batches.

## 6. Conclusion

In this paper has been proposed a corrected inertial tolerancing method, which aims to guarantee the conformity of the assembly resultant in a tolerance interval thanks to a *Cpk* index. The *ICC* coefficient value to take into account for the tolerances calculation is found regarding the number of components in the assembly, and the minimum value of the *Cpk* index to guarantee on the assembly resultant.

The application of this tolerancing method allows to apply the inertial tolerancing method but with a link to the traditional tolerancing. That is to say that this method allows to guarantee a *Cpk* index on the assembly resultant while applying the inertial tolerancing based on a statistical approach.

## 7. Bibliography

Pillet M., Bernard F., Avrillon L, « Le tolérancement inertiel, une autre façon d'intégrer l'aspect combinatoire dans les processus assemblés », *Congrès CPI* 2001 – Fès Maroc.

Graves S., « Tolerance Analysis Taylored to your organization », *journal of Quality technology*, Vol 33, n°3, 2001.

Schewart W.A. *Economic Control of Quality of Manufactured Product.* Van Nostrand, New York, 1931.

Evans D.H. « Statistical tolerancing: the state of the art », *journal of quality technology*, Vol 7 n°1, 1975.

Chase K.W., Prakinson A.R. « A survey of research in the application of tolerance analysis to the design of mechanical assemblies », *Research in Engineering design* 3, 1991.

Graves S. « How to reduce costs using a tolerance analysis formula tailored to your organization », *CQPI Report*, n° 157, 1997.

Grave S., Bisgaard S. « Five ways statistical tolerancing can fail and what to do about them », *Quality engineering*, vol 13, 1997, p.85-93.

Pillet M. « Inertial Tolerancing in the case of assembled products », *IDMME* 2002